# Unexpected structural and magnetic depth dependence of YIG thin films

**J.F.K. Cooper, C.J. Kinane, S. Langridge**

ISIS Neutron and Muon Source, Rutherford Appleton Laboratory, Harwell Campus, Didcot, OX11 0QX

**M. Ali, B.J. Hickey**

Condensed Matter group, School of Physics and Astronomy, E.C. Stoner Laboratory, University of Leeds, LS2 9JT

**T. Niizeki**

WPI Advanced Institute for Materials Research, Tohoku University, Sendai 980-8577, Japan

**K. Uchida**

National Institute for Materials Science, Tsukuba 305-0047, Japan

Institute for Materials Research, Tohoku University, Sendai 980-8577, Japan

Center for Spintronics Research Network, Tohoku University, Sendai 980-8577, Japan

PRESTO, Japan Science and Technology Agency, Saitama 332-0012, Japan

**E. Saitoh**

WPI Advanced Institute for Materials Research, Tohoku University, Sendai 980-8577, Japan

Institute for Materials Research, Tohoku University, Sendai 980-8577, Japan

Center for Spintronics Research Network, Tohoku University, Sendai 980-8577, Japan

WPI Advanced Institute for Materials Research, Tohoku University, Sendai 980-8577, Japan

Advanced Science Research Center, Japan Atomic Energy Agency, Tokai 319-1195, Japan

**H. Ambaye**

Neutron Sciences Directorate, Oak Ridge National Laboratory, Oak Ridge, Tennessee 37831, USA

**A. Glavic**

Laboratory for Neutron Scattering and Imaging, Paul Scherrer Institut, Villigen PSI, Switzerland

Neutron Sciences Directorate, Oak Ridge National Laboratory, Oak Ridge, Tennessee 37831, USA

PACS: 75.70.-I, 75.47.Lx, 68.55.aj

## Abstract

We report measurements on yttrium iron garnet (YIG) thin films grown on both gadolinium gallium garnet (GGG) and yttrium aluminium garnet (YAG) substrates, with and without thin Pt top layers. We provide three principal results: the observation of an interfacial region at the Pt/YIG interface, we place a limit on the induced magnetism of the Pt layer and confirm the existence of an interfacial layer at the GGG/YIG interface. Polarised neutron reflectometry (PNR) was used to give depth dependence of both the structure and magnetism of these structures. We find that a thin film of YIG on GGG is best described by three distinct layers: an interfacial layer near the GGG, around 5 nm thick and non-magnetic, a magnetic 'bulk' phase, and a non-magnetic and compositionally distinct thin layer near the surface. We theorise that the bottom layer, which is independent of the film thickness, is caused by Gd diffusion. The top layer is likely to be extremely important in inverse spin Hall effect measurements, and is most likely $Y_2O_3$ or very similar. Magnetic sensitivity in the PNR to any induced moment in the Pt is increased by the existence of the $Y_2O_3$ layer; any moment is found to be less than 0.02 uB/atom.

## Introduction

Yttrium iron garnet (YIG) has long been known to be a ferrimagnetic insulator and is used widely as a tuneable microwave filter or, when doped with other rare earth elements, for a variety of optical and magneto-optical applications. However, since the discovery of the spin Seebeck effect in insulators[1,2], YIG, particularly when grown on gadolinium gallium garnet (GGG), has become the model system for investigating the physics of the spin Seebeck effect. The spin Seebeck effect combines two future technologies: the pure spin currents of spintronics promise to eliminate Joule heating in computing and many other industries[3–5], whereas energy recovery materials seek to harvest waste heat and movement to reduce energy losses, either actively[6–8], or passively from the conventional Seebeck effect[9] or otherwise[10]. The combination of these into a single material provides a great opportunity for energy efficiency and a new generation of future devices.

It is therefore extremely important that both the interfacial physics of GGG/YIG and the physical system, with all possible imperfections, are well understood. Much work has been done on the theoretical understanding of the spin Seebeck effect[11–14], with a general consensus that the effect is magnon driven, with non-equilibrium phonons also playing a role. This paper seeks to explore the material science aspect of the GGG/YIG system, and to understand the effects of interfacial structure on high quality epitaxial films.

Thin films of YIG, with different annealing times, were grown on GGG by sputtering and these films were characterised using polarised neutron reflectivity (PNR) to extract a magnetic depth profile, as well as x-ray reflectivity (XRR) and magnetometry. Films were measured both with and without thin Pt layers on top of the YIG; these layers are conventionally used for inverse spin Hall effect (ISHE) measurements, to quantify the strength of the spin Seebeck effect. Additional films were grown on yttrium aluminium garnet (YAG) in order to investigate the effects of the substrate on the films.

## Methods

Samples were grown in both Leeds and Tohoku Universities with common growth methodologies with the exception of the annealing time. The Leeds samples were sputtered in a RF magnetron sputter chamber with a base pressure of $2\times10^{-8}$ Torr, with oxygen and argon flow rates of 1.2 and 22.4 sccm respectively. They were deposited onto either GGG or YAG substrates, 1" in diameter. The samples were then removed from the vacuum and annealed in air at 850 $^o$C for 2 hours. They were then sputtered with a thin layer of Pt ~ 27 Å thick, a typical thickness for ISHE measurements. The films grown in Tohoku were prepared according to the methods detailed in the work by Lustikova *et al*[15], where the same annealing temperature was used, but for 24 hours instead of 2.

A total of eight samples were measured: six from Leeds, four grown on GGG substrates and two on YAG, and two from Tohoku, both on GGG. The Leeds films on GGG were either 300 Å or 800 Å thick with and without a thin Pt layer on top. Both of the YAG films were 800 Å, and one had a Pt top layer. The Tohoku films were both around 1500 Å thick, one with a Pt layer (135 Å thick) on top and one without.

PNR measurements record the neutron reflectivity as a function of the neutron's wave-vector transfer and spin eigenstate. Modelling of the resultant data allows the scattering length density (SLD) to be extracted and provides a quantitative description of the depth dependent structural and magnetic profile[16].

PNR measurements were taken on the Polref and Offspec[17] beamlines at the ISIS neutron and muon source, as well as the Magnetism Reflectometer beamline at the Spallation Neutron Source, in Oak Ridge. X-ray measurements and magnetometry was carried out in the R53 Characterisation lab at ISIS.

Fitting to both the neutron reflectivity and the x-ray reflectivity data was performed in the GenX fitting package[18]. The Pt cap layers thickness were determined by XRR since the scattering contrast between Pt and YIG or GGG is very good for x-rays and reduced for neutrons, whereas the contrast between YIG and GGG is poor for x-rays and good for neutrons. Magnetometry was performed using a Durham Magneto Optics NanoMOKE3 and showed that all of the films presented here had a coercivity of <5 Oe and generally ~1-2 Oe, indicating high quality YIG.

## Results and Discussion

Figure 1 presents the polarised neutron reflectivity and resultant SLD for the YIG layer on GGG and YAG substrates. From the nominal structure of the sputtered samples the neutron reflectivity can be calculated. This simple model does not accurately describe the observed data. To describe the GGG system an additional interfacial YIG-like layer was required, see layer (a) in Figure 1. This layer was either non-magnetic, or had a very small moment (~<0.1 $\mu_B$ /unit cell), and was ~50 Å thick, irrespective of the total YIG film thickness. This layer was not present in the films grown on a YAG substrate. The large roughness of the interface between the GGG and the YIG in the models suggests that a diffusion process created this layer. This layer was not formed at the interface with the YAG substrates indicating that this process must be either Ga or Gd diffusion. A recent temperature dependent study of this interface also using neutrons showed that it is Gd[19].

Beyond the initial 50 Å non-magnetic layer, the structural and magnetic properties of the sputtered YIG films of differing thickness were very similar and did not have any thickness dependences. This was true of the films grown on both GGG and YAG, meaning that the effects of the substrate, for these at least, are minimal beyond its ability to diffuse during annealing.

From the magnetic SLD the moment of the bulk YIG (away from both interfaces) was found to be consistently 3.8(1) $\mu_B$ / unit cell, this value did not depend on the film thickness. This value compares well with literature values[20] at room temperature, though very slightly higher.

The measurements on the YIG films with extended annealing times were less conclusive. Models both with and without the non-magnetic layer at the GGG interface gave similar fits to the PNR data. These films were significantly thicker than the films with a 2 hr anneal, ~1500 Å, and as such the sensitivity to the GGG/YIG interface is reduced. As a results it is not possible to conclusively identify the presence of an interfacial layer. Since the annealing procedures for both sets of films are very similar we can assume that the Gd diffusion will also be similar, and a common feature of the GGG/YIG interface.

In addition to the substrate interface layer, an additional layer was discovered for all of the samples, labelled as layer (b) in Figure 1. This layer is around 15(5) Å in thickness, with little variation, across all samples measured. This layer was distinct from both the Pt and the YIG, as it had a markedly lower scattering length density than either of them. Figure 2 shows datasets for both thin Leeds (300 Å) and thick Tohoku (1500 Å) YIG on GGG, with the best fit to the data. Since the SLD of YIG and Pt is similar, the low frequency oscillations in both datasets results from the low SLD layers contrast between the Pt and the YIG. Analysing x-ray reflectivity curves of the same samples, with and without the Pt cap also require this layer. Examining the two scattering length densities (neutrons and x-rays) of the layer involved we can elucidate its composition. Pt alloys would generate a strong x-ray contrast, and the layer would not appear in uncapped samples and can therefore be ruled out. Both iron and all forms of its oxide have too large an SLD for neutrons so it can be ruled out. This leaves yttrium based compounds: pure Y, $Y_2O_3$, (yttria) and YN (which is possible, though unlikely, due to the annealing of the sample in air). The x-ray SLD of $Y_2O_3$ is a close match with the SLD of the layer required for a good fit, as shown in Figure 3. In addition to the matching SLD, we remark that $Y_3Fe_5O_{12}$ has an oxidation state of +3 for Y, which is the same as $Y_2O_3$ and both have similar oxygen co-ordination. The Y-O bond length in $Y_2O_3$ is between 2.225 and 2.323 Å[21], which represents a slight contraction with respect to the bond length in YIG, which is between 2.37 and 2.43 Å[22].

The magnetic signal from the YIG decays across this layer (see Figure 3) and as it is likely that the layer is predominantly yttria (a non-magnetic insulator), the electrical resistivity at some point in the film becomes that of the Pt. This means that any ISHE effects are likely sampling a lot more of the yttria, than the YIG. Several studies have investigated the influence of the interface quality[23–25] and have determined that, as might be expected, a high quality interface yields better ISHE results. Qiu et al.[24] found that, the ISHE voltage varied from around 3.5 µV/K for a minimal interface yttria region, to nearly 0 for regions over 7 nm thick. They also found that, at least for samples grown by liquid phase epitaxy, optimisation of after growth annealing could minimise this layer's formation. This interface has also been investigated by Song et al.[26] using electron microscopy, though they attribute the layer to being oxygen deficient iron (whose magnetic moment is then reduced).

Knowledge of the existence and the information about the nature of this layer extracted here gives an opportunity to eliminate it in all cases; either by appropriate annealing, or selective etching.

As a result of the low SLD of this layer and its contrast to the Pt layer, we have gained unusual sensitivity to any induced magnetism in the Pt layer. Proximity effects in the Pt have been widely studied, with many works extracting the origins of the observed voltages from the iSHE[27–29]. Theoretical studies such as Liang et al.[30] show how a non-magnetic, or reduced magnetic layer would be important in proximity effects in this system. Figure 4 shows a portion of the spin asymmetry (difference in reflectivity of the two spin states normalised by their sum) from GGG/YIG(8000)/Pt(27), at high momentum transfer. The spin asymmetry is sensitive to the magnetism, and is, to a first approximation, independent of the exact structure. The reflectivity can be approximated as the Fourier transform of the layer structure, so at high Q we are sensitive to thinner layers, e.g. the top Pt layer.

The best model line is shown in grey and is clearly a good fit, even by this, more highly processed, measure. In addition to the best fit in Figure 4 (shown in grey) are two models (blue and red), with the same structural parameters and bulk YIG magnetism, but with an induced moment added to the Pt of ±0.05 $\mu_B$/Pt atom to show how this would affect the fit. From the deviation of these models from the data, we can clearly see that the average magnitude of any induced moment in the Pt is certainly less than 0.05 $\mu_B$/Pt atom, and likely less than 0.02 $\mu_B$/Pt atom, within a 1σ error bound. This result is in line with previous experiments[31] which specifically tried to measure an induced moment in Pt, albeit with the YIG grown by a different method. It is worth noting here that it may still be possible that a more magnetic sub-region of the Pt may exist, since PNR cannot probe infinitely thin layers. However, the magnitude of the total magnetism would still have to remain below our upper bound, e.g. if half was non-magnetic and half was polarised, then it would have to be below 0.04 $\mu_B$/Pt atom, etc.

# Conclusion

We have used polarised neutron reflectivity to determine the magnetic depth dependence of yttrium iron garnet thin films grown on gadolinium gallium garnet and yttrium aluminium garnet substrates, with and without a Pt layer on the surface. It was found that if the YIG is grown on a GGG substrate there can be a ~50 Å non-magnetic layer at the substrate interface, this does not depend on the YIG film thickness. This is likely to be caused by Gd diffusion during annealing, since this layer does not appear when the YIG is grown on YAG, this is in line with recent investigations[19]. The effect of growing YIG on YAG, other than the absence of this interface layer was minimal, with roughnesses and magnetic moments extremely similar to those grown on GGG.

We also see an additional layer at the YIG/Pt interface, roughly 15 Å for all samples measured; further investigation and cross referencing with x-ray measurements identifies this layer as $Y_2O_3$. While the existence of this (non-magnetic and usually insulating) layer may have large repercussions for the interpretation of ISHE measurements on this model system, knowledge of its existence and composition means it may be possible to eliminate it.

Our measurements also give us unusual sensitivity to any induced magnetism in the Pt layer, and allow us to give an upper bound on the magnitude of the moment of ±0.02 $\mu_B$/Pt atom.

## Acknowledgements

The neutron work in this paper was performed at both the Spallation Neutron Source in the Oak Ridge National Laboratory (IPTS-13192), USA, and at the ISIS Pulsed Neutron and Muon Source, which were supported by a beamtime allocation from the Science and Technology Facilities Council (RB1410610 and RB1510146). We would like to thank the sample environment support staff at both facilities for their help with the experiments.

This work is partially supported by PRESTO "Phase Interfaces for Highly Efficient Energy Utilization" and ERATO "Spin Quantum Rectification Project" from JST, Japan, and by Grant-in-Aid for Scientific Research (A) (No. JP15H02012) and Grant-in-Aid for Scientific Research on Innovative Area "Nano Spin Conversion Science" (No. JP26103005) from JSPS KAKENHI, Japan.

## References

[1] K. Uchida, J. Xiao, H. Adachi, J. Ohe, S. Takahashi, J. Ieda, T. Ota, Y. Kajiwara, H. Umezawa, H. Kawai, G.E.W. Bauer, S. Maekawa, and E. Saitoh, Nat. Mater. **9**, 894 (2010).

[2] K. Uchida, H. Adachi, T. Ota, H. Nakayama, S. Maekawa, and E. Saitoh, Appl. Phys. Lett. **97**, 172505 (2010).

[3] I. Žutić and S. Das Sarma, Rev. Mod. Phys. **76**, 323 (2004).

[4] D.D. Awschalom and M.E. Flatté, Nat. Phys. **3**, 153 (2007).

[5] S.A. Wolf, D.D. Awschalom, R.A. Buhrman, J.M. Daughton, S. von Molnár, M.L. Roukes, A.Y. Chtchelkanova, and D.M. Treger, Science **294**, 1488 (2001).

[6] R.J.M. Vullers, R. van Schaijk, I. Doms, C. Van Hoof, and R. Mertens, Solid. State. Electron. **53**, 684 (2009).

[7] D. Guyomar, A. Badel, E. Lefeuvre, and C. Richard, IEEE Trans. Ultrason. Ferroelectr. Freq. Control **52**, 584 (2005).

[8] S.P. Beeby, M.J. Tudor, and N.M. White, Meas. Sci. Technol. **17**, R175 (2006).

[9] T.J. Abraham, D.R. MacFarlane, and J.M. Pringle, Energy Environ. Sci. **6**, 2639 (2013).

[10] D.L. Andrews, in *Int. Symp. Opt. Sci. Technol.*, edited by A. Lakhtakia, G. Dewar, and M.W. McCall (International Society for Optics and Photonics, 2002), pp. 181–190.

[11] J. Xiao, G.E.W. Bauer, K. Uchida, E. Saitoh, and S. Maekawa, Phys. Rev. B **81**, 214418 (2010).

[12] H. Adachi, K. Uchida, E. Saitoh, and S. Maekawa, Rep. Prog. Phys. **76**, 36501 (2013).

[13] H. Adachi, J. Ohe, S. Takahashi, and S. Maekawa, Phys. Rev. B **83**, 94410 (2011).

[14] S.M. Rezende, R.L. Rodríguez-Suárez, R.O. Cunha, A.R. Rodrigues, F.L.A. Machado, G.A. Fonseca Guerra, J.C. Lopez Ortiz, and A. Azevedo, Phys. Rev. B **89**, 14416 (2014).

[15] J. Lustikova, Y. Shiomi, Z. Qiu, T. Kikkawa, R. Iguchi, K. Uchida, and E. Saitoh, J. Appl. Phys. **116**, 153902 (2014).

[16] B.P. Toperverg and H. Zabel, in *Exp. Methods Phys. Sci.*, 48th ed. (2015), pp. 339–434.

[17] J.R.P. Webster, S. Langridge, R.M. Dalgliesh, and T.R. Charlton, Eur. Phys. J. Plus **126**, 112 (2011).

[18] M. Björck and G. Andersson, J. Appl. Crystallogr. **40**, 1174 (2007).

[19] A. Mitra, O. Cespedes, Q. Ramasse, M. Ali, S. Marmion, M. Ward, R.M.D. Brydson, C.J. Kinane, J.F.K. Cooper, S. Langridge, and B.J. Hickey, Sci. Rep. **submitted**, (2017).

[20] A. Bouguerra, G. Fillion, E.K. Hlil, and P. Wolfers, J. Alloys Compd. **442**, 231 (2007).

[21] J. Zhang, F. Paumier, T. Höche, F. Heyroth, F. Syrowatka, R.J. Gaboriaud, and H.S. Leipner, Thin Solid Films **496**, 266 (2006).

[22] S. Geller and M.A. Gilleo, J. Phys. Chem. Solids **3**, 30 (1957).

[23] Z. Qiu, K. Ando, K. Uchida, Y. Kajiwara, R. Takahashi, H. Nakayama, T. An, Y. Fujikawa, and E. Saitoh, Appl. Phys. Lett. **103**, 92404 (2013).

[24] Z. Qiu, D. Hou, K. Uchida, and E. Saitoh, J. Phys. D. Appl. Phys. **48**, 164013 (2015).

[25] S. Vélez, A. Bedoya-Pinto, W. Yan, L.E. Hueso, and F. Casanova, (2016).

[26] D. Song, L. Ma, S. Zhou, and J. Zhu, Appl. Phys. Lett. **107**, 42401 (2015).

[27] S.Y. Huang, X. Fan, D. Qu, Y.P. Chen, W.G. Wang, J. Wu, T.Y. Chen, J.Q. Xiao, and C.L. Chien, Phys. Rev. Lett. **109**, 107204 (2012).

[28] T. Kikkawa, K. Uchida, Y. Shiomi, Z. Qiu, D. Hou, D. Tian, H. Nakayama, X.-F. Jin, and E. Saitoh, Phys. Rev. Lett. **110**, 67207 (2013).

[29] T. Kikkawa, K. Uchida, S. Daimon, Y. Shiomi, H. Adachi, Z. Qiu, D. Hou, X.-F. Jin, S. Maekawa, and E. Saitoh, Phys. Rev. B **88**, 214403 (2013).

[30] X. Liang, Y. Zhu, B. Peng, L. Deng, J. Xie, H. Lu, M. Wu, and L. Bi, ACS Appl. Mater. Interfaces **8**, 8175 (2016).

[31] S. Geprägs, S. Meyer, S. Altmannshofer, M. Opel, F. Wilhelm, A. Rogalev, R. Gross, and S.T.B. Goennenwein, Appl. Phys. Lett. **101**, 262407 (2012).

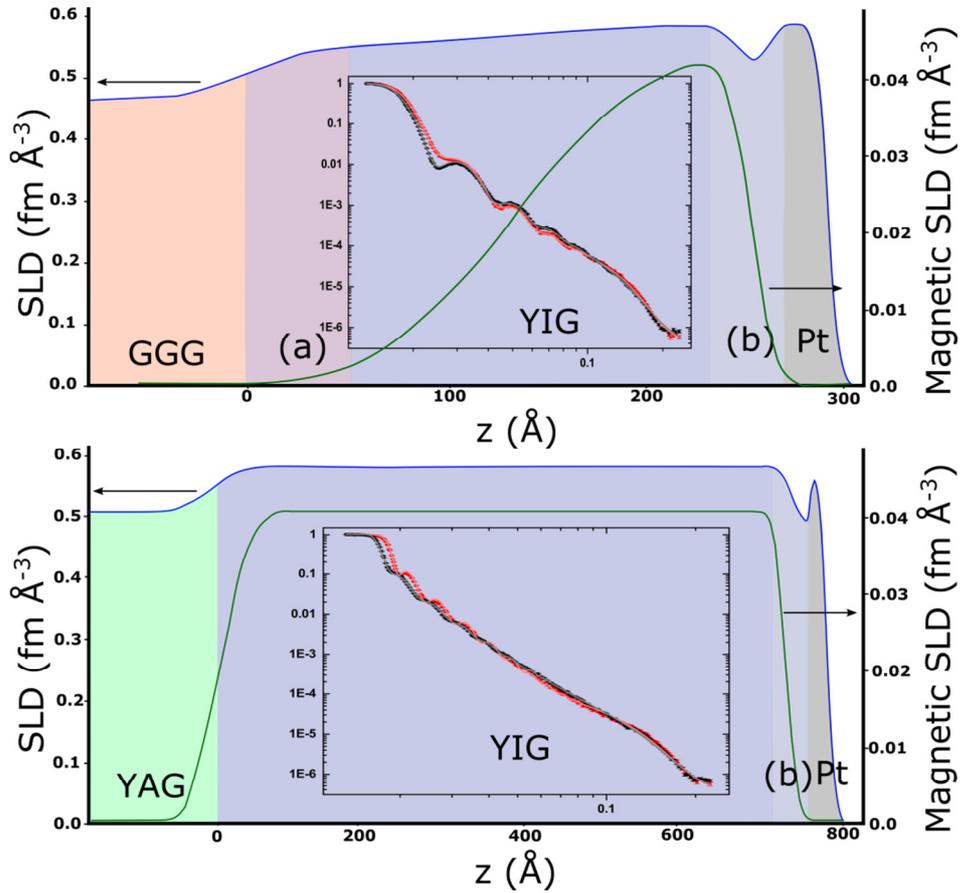

**Figure 1** Structural (blue) and magnetic (green) scattering length densities of fitted models for a YIG/Pt bilayer grown on a GGG substrate (top) and YAG substrate (bottom). The top shows a thin ~300 Å YIG layer, and the bottom shows a thick ~800 Å YIG layer, however, the data are representative of all the samples grown on each substrate irrespective of YIG thickness. The GGG substrate has two extra layers: (a), at the substrate interface, and (b), at the YIG/Pt interface. (a) is non-magnetic at room temperature and around 50 Å thick irrespective of the thickness of the YIG layer. This layer is not present at the YAG/YIG interface. The nature of layer (b) discussed later and is absolutely required for a good fit in all samples with a Pt layer.

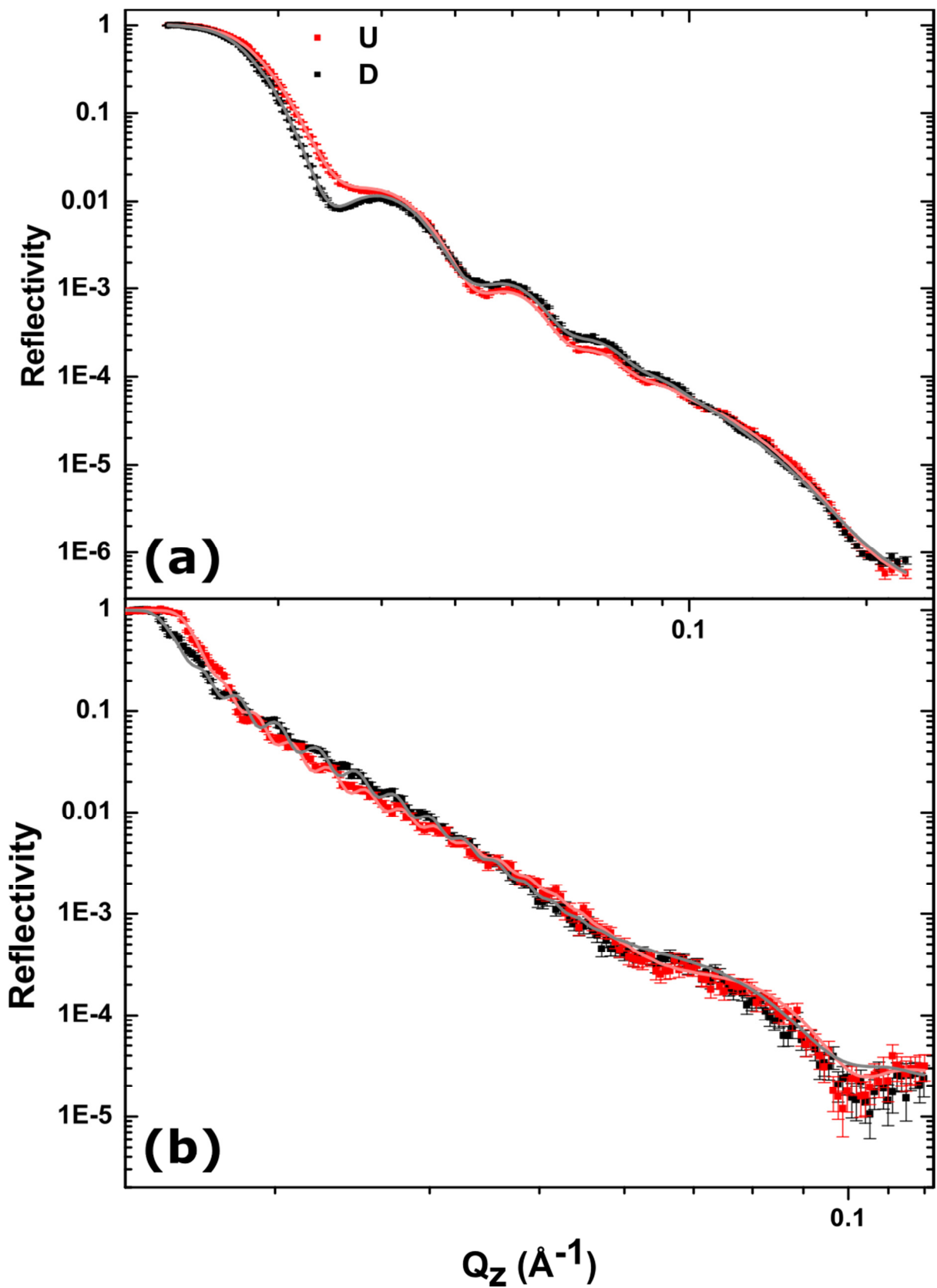

**Figure 2** Polarised Neutron Reflectivity data (points) and modelled fit (line) of thin (~300 Å) GGG/YIG/Pt from Leeds, (a), and thicker (~1500 A) Tohoku, (b), samples, both of which have Pt top layers. The low frequency oscillations (which is the majority of the curve in (a) since it has a thinner Pt layer) are visible due to the low scattering length density layer between the YIG and the Pt and is required in models for all samples with Pt on to get a reasonable fit.

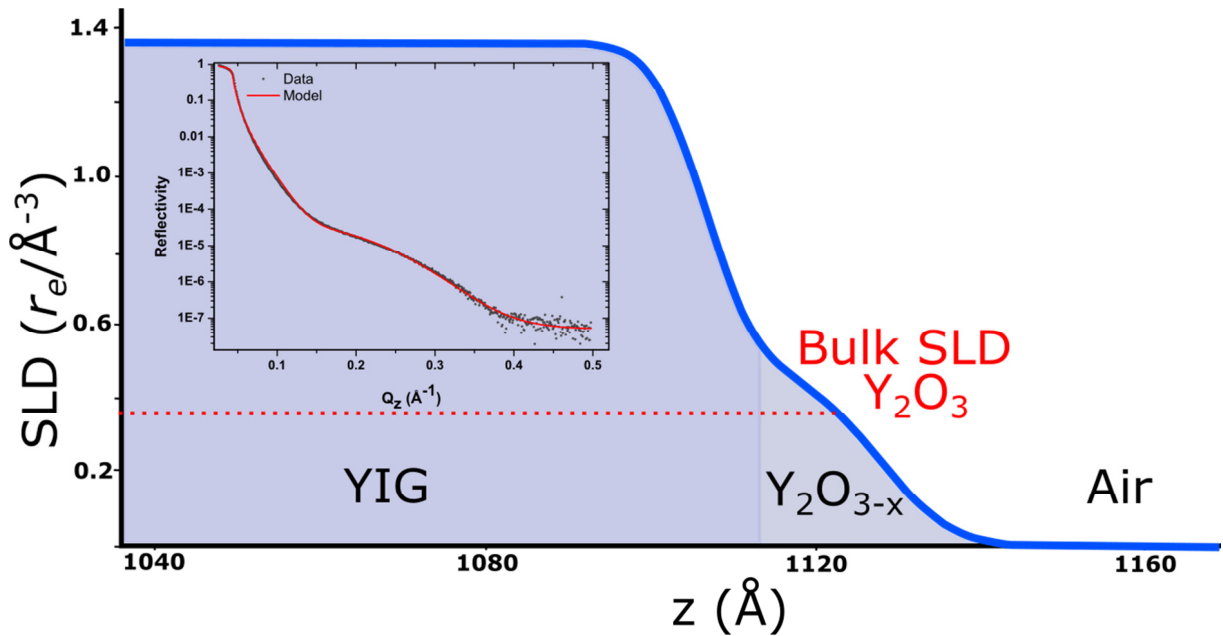

**Figure 3** X-ray scattering length density of the interface between YIG and air, with x-ray reflectivity data and fit inset. The subtle step in the SLD is required in order to correctly model the slow oscillation in the reflectivity data. By using both the x-ray and neutron scattering length densities we can deduce that the top layer in this, and all other samples measured in this study, is extremely likely to be yttria ($Y_2O_3$), whose bulk SLD is indicated by the dotted red line.

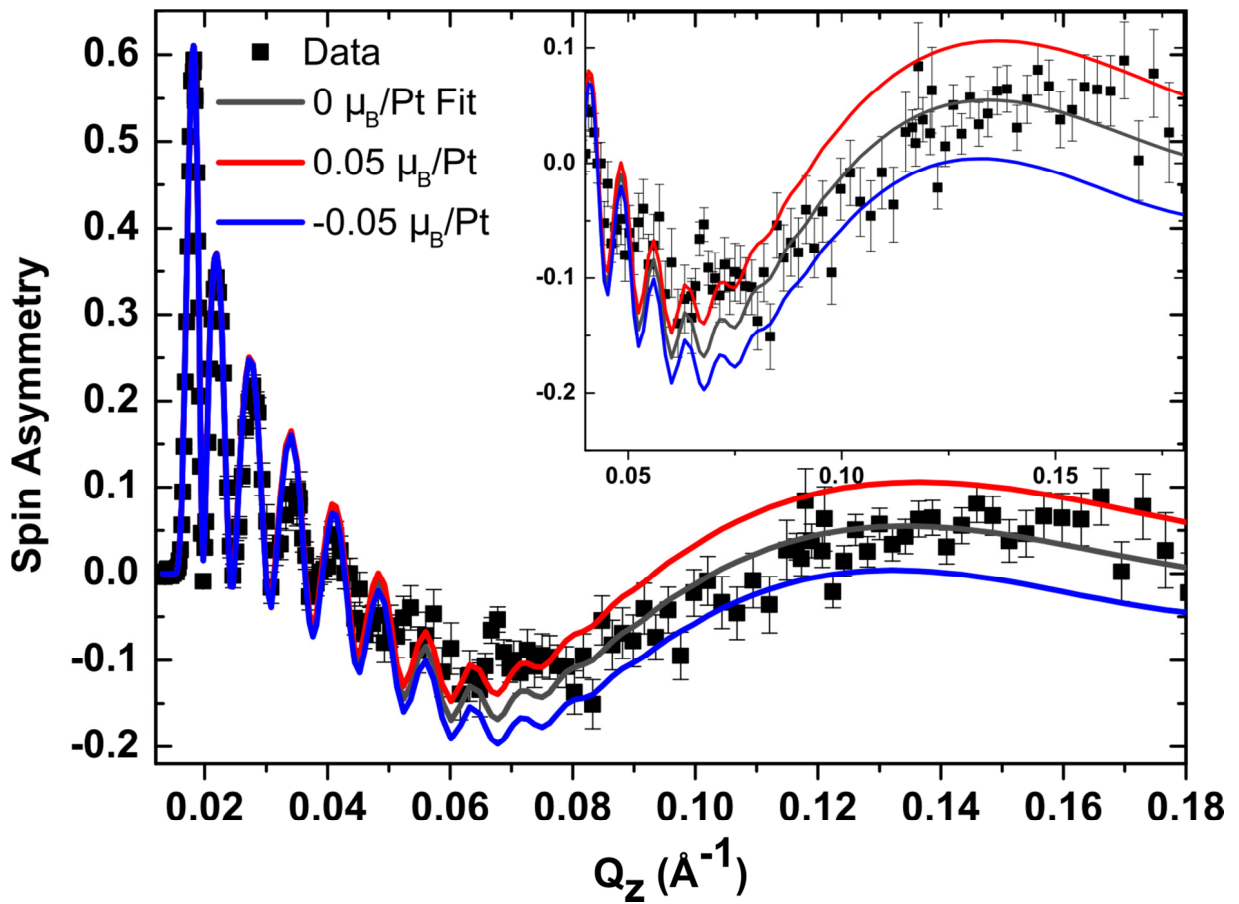

**Figure 4** A selected portion of the spin asymmetry of a thick (~800 Å) YAG/YIG/Pt structure, with a zoom inset. The higher Q range is where we are sensitive to any induced magnetism in the Pt layer. The grey curve shows the asymmetry produced by the optimal fit with no induced magnetism in the Pt. The red and blue curves show the same model with +0.05 $\mu_B$/Pt atom and -0.05 $\mu_B$/Pt atom of induced magnetism. A similar result is found for all samples measured, irrespective of YIG thickness or substrate.